\RequirePackage{lineno}
\documentclass[10pt]{IEEEtran} 
\usepackage{epsfig,amsmath,amsfonts,cite}
\usepackage{threeparttable}
\usepackage{lipsum}
\usepackage{color}
\usepackage{epstopdf}
\usepackage{graphicx}
\usepackage{lipsum}
\usepackage{algorithm,algorithmic}
\usepackage{epsfig,amsmath,amsfonts,cite}
\usepackage{booktabs}
\newcommand{\thickhline}{\noalign{\hrule height 1.0pt}}

\newcounter{MYtempeqncnt}
\usepackage{multirow}
\usepackage{multicol}
\ifCLASSINFOpdf
 
\else
\fi

\begin{document}

\title{Calculation of Generalized Polynomial-Chaos Basis Functions and Gauss Quadrature Rules in Hierarchical Uncertainty Quantification}

\author{Zheng Zhang,~\IEEEmembership{Student Member,~IEEE}, Tarek A. El-Moselhy, Ibrahim (Abe) M. Elfadel,~\IEEEmembership{Senior Member,~IEEE}, and~Luca~Daniel,~\IEEEmembership{Member,~IEEE}

\thanks{This work was funded by the MIT-SkolTech program and by the Cooperative Agreement between the Masdar Institute of Science and Technology, Abu Dhabi, UAE and the Massachusetts Institute of Technology (MIT), Cambridge, MA, USA (Reference No. 196F/002/707/102f/70/9374).}      
\thanks{Z. Zhang and L. Daniel are with the Department of Electrical Engineering and Computer Science, Massachusetts Institute of Technology (MIT), Cambridge, MA 02139, USA (e-mail: z\_zhang@mit.edu, luca@mit.edu).}
\thanks{T. El-Moselhy was with the Department of Aeronautics and Astronautics, Massachusetts Institute of Technology (MIT), Cambridge, MA 02139, USA. He is now with the D. E. Shaw Group, 1166 Avenue of the Americas, New York, NY 10036, USA (e-mail: tarek.moselhy@gmail.com).}
\thanks{I. Elfadel is with the Masdar Institute of Science and Technology, United Arab Emirates (e-mail: ielfadel@masdar.ac.ae).}
}

\markboth{IEEE TRANSACTIONS ON COMPUTER-AIDED DESIGN OF INTEGRATED CIRCUITS AND SYSTEMS, ~Vol. ~XX, No.~XX,~XX~2014}{ZHANG \MakeLowercase{\textit{et al.}}: Density Estimation and Parameter Reduction for Statistical Circuit}

\maketitle

\begin{abstract}
Stochastic spectral methods are efficient techniques for uncertainty quantification. Recently they have shown excellent performance in the statistical analysis of integrated circuits. In stochastic spectral methods, one needs to determine a set of orthonormal polynomials and a proper numerical quadrature rule. The former are used as the basis functions in a generalized polynomial chaos expansion. The latter is used to compute the integrals involved in stochastic spectral methods. Obtaining such information requires knowing the density function of the random input {\it a-priori}. However, individual system components are often described by surrogate models rather than density functions. In order to apply stochastic spectral methods in hierarchical uncertainty quantification, we first propose to construct physically consistent closed-form density functions by two monotone interpolation schemes. Then, by exploiting the special forms of the obtained density functions, we determine the generalized polynomial-chaos basis functions and the Gauss quadrature rules that are required by a stochastic spectral simulator. The effectiveness of our proposed algorithm is verified by both synthetic and practical circuit examples.
\end{abstract}


\IEEEpeerreviewmaketitle

\section{Introduction}

\IEEEPARstart{D}{ue} to significant manufacturing process variation, it has become necessary to develop efficient uncertainty quantification tools for the fast statistical analysis of electronic circuits and systems~\cite{zzhang:tcad2013,zzhang:tcas2_2013,zzhang:iccad_2013,Pulch:2011,Pulch:2011_1,SingheeR09,SingheeR10,Felt:1996,Wang:2004,Vrudhula:2006,sMOR2012,Tarek_ISQED:11,Tarek_DAC:08,TarekDATE2010,TarekDAC2008,Hchang:2005,Hchang:2003,Jaskirat:2006}. Monte Carlo simulators~\cite{SingheeR09,SingheeR10,Felt:1996} have been utilized in statistical circuit analysis for decades. Recently, stochastic spectral methods~\cite{UQ:book,col:2005,Ivo:2007,Nobile:2008,Nobile:2008_2,sfem} have emerged as a promising technique for the uncertainty quantification of integrated circuits~\cite{zzhang:tcad2013,zzhang:tcas2_2013,zzhang:iccad_2013,Pulch:2011,Pulch:2011_1}. Such methods approximate the stochastic solution by a truncated generalized polynomial chaos expansion~\cite{gPC2002,gPC2003,xiu2009}, which converges much faster than Monte Carlo when the parameter dimensionality is not high.

\begin{figure}[t]
	\centering
		\includegraphics[width=3.3in]{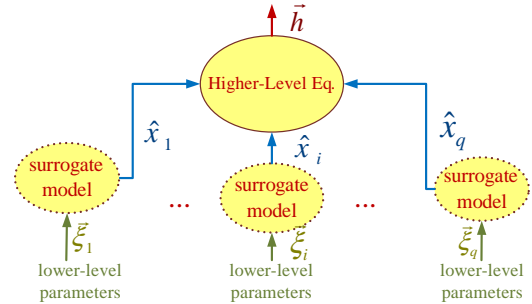} 
\caption{Demonstration of hierarchical uncertainty quantification.}
	\label{fig:hierarchical}
\end{figure}

This work is motivated by the need for hierarchical uncertainty quantification based on stochastic spectral methods. Consider Fig.~\ref{fig:hierarchical} that demonstrates the uncertainty quantification of a complex system. In this system, there exist $q$ readily obtained surrogate models:
\begin{equation}
\label{surrogate_i}
\hat x_i=f_i(\vec \xi_i), \; {\rm with} \;\vec \xi_i\in \mathbb{R}^{d_i}, \; i=1,\cdots, q
\end{equation}
where $\hat x_i$ is a variable dependent on multiple lower-level random parameters. In transistor-level simulation, $\hat x_i$ is a device-level parameter (e.g., threshold voltage of a transistor) influenced by some geometric and process variations. In a statistical behavior-level simulator~\cite{Felt:1996}, $\hat x_i$ is the performance metric of a small circuit block (e.g., the frequency of a voltage-controlled oscillator) affected by some device-level parameters $\vec \xi_i$. Typical surrogate models include linear (quadratic) response surface models~\cite{TCAD2006,xli2004:Apex,Apex,PEM,CMichael:1992,Felt:1996}, truncated generalized polynomial chaos representations~\cite{zzhang:tcad2013,zzhang:tcas2_2013}, smooth or non-smooth functions, stochastic reduced-order models~\cite{sMOR2012,TarekDATE2010,Frangos:2010}, and some numerical packages that can rapidly evaluate $f_i(\vec \xi_i)$ (e.g., computer codes that implement a compact statistical device model). By solving a system-level equation, the output $\vec h$ can be obtained.

Instead of simulating the whole system starting from the bottom-level random parameters $\vec \xi_i$'s, uncertainty quantification can be performed in a hierarchical way. By treating $\hat x_i$'s as the random inputs for the higher-level equation, the output $\vec h$ can be computed more efficiently. This treatment can dramatically reduce the parameter dimensionality and problem size. Related work in this direction includes the statistical analysis of phase-lock loops~\cite{Felt:1996} and the statistical timing analysis of digital VLSI~\cite{Hchang:2005,Hchang:2003,Jaskirat:2006}. In all existing works, Monte Carlo was utilized to perform the higher-level simulation. 

When $\hat x_i$'s are mutually independent, it is possible to further speed up the hierarchical uncertainty quantification flow by using stochastic spectral methods~\cite{zzhang:tcad2013,zzhang:tcas2_2013,zzhang:iccad_2013,Pulch:2011,Pulch:2011_1,UQ:book,col:2005,Ivo:2007,Nobile:2008,Nobile:2008_2,sfem}. In this case, high-accuracy results may be obtained by fast simulation if $\vec h$ smoothly depends on $\hat x_i$'s (even if $f_i(\vec \xi_i)$ is non-smooth). In order to apply available stochastic spectral methods, we need to determine a set of orthonormal polynomials as the basis functions for generalized polynomial chaos expansions. Sometimes we also need a proper numerical quadrature rule (such as Monte Carlo, Gauss quadrature, etc.)~\cite{zzhang:iccad_2013}. In this paper, we will consider Gauss quadrature since it is widely used in stochastic spectral methods. Both tasks require the probability density function of each random input $\hat x_i$. Existing techniques typically assume that the random inputs have some well-known distributions (e.g., Gaussian, uniform, Gamma and Beta distributions), and make use of available quadrature rules and orthogonal polynomials (e.g., Hermite, Legendre, Laguerre and Jacobi polynomials)~\cite{zzhang:tcad2013,zzhang:tcas2_2013,zzhang:iccad_2013,Pulch:2011,Pulch:2011_1,col:2005,Ivo:2007,Nobile:2008,Nobile:2008_2,sfem,gPC2002,gPC2003,xiu2009}. This assumption obviously does not hold in our case: the probability density function of $\hat x_i$ is not readily available from its surrogate model.

Therefore, a question is how to determine the generalized polynomial-chaos basis functions and Gauss quadrature rules from a general surrogate model. This paper aims to partly answer this key question in hierarchical uncertainty quantification. Our method is based on the ideas of changing variables and monotone interpolation~\cite{Fritsch:1980,Hyman:1983,Gregory:1985,Gregory:1982}. Specifically, we represent the random input as a linear function of a new parameter, and treat such parameter as a new random input. Using two monotone interpolation schemes, physically consistent closed-form cumulative density functions and probability density functions can be constructed for the new random input. Due to the special forms of the obtained density functions, we can easily determine a proper Gauss quadrature rule and the basis functions for a generalized polynomial chaos expansion.

We focus on the general framework and verify our method by using both synthetic and performance-level circuit surrogate models. Our method can be employed to handle a wide variety of surrogate models, including device-level models for SPICE-level simulators~\cite{zzhang:tcad2013,zzhang:tcas2_2013,zzhang:iccad_2013}, circuit-level performance models for behavior-level simulation~\cite{Felt:1996}, as well as gate-level statistical models for the timing analysis of digital VLSI~\cite{Hchang:2005,Hchang:2003,Jaskirat:2006}. In this paper we will focus only on the derivation of the basis functions and Gauss quadrature rules and refer the reader interested to the extensive literature on how to use them in a stochastic spectral simulator (see~\cite{zzhang:tcad2013,zzhang:tcas2_2013,zzhang:iccad_2013,Pulch:2011,Pulch:2011_1,col:2005,Ivo:2007,Nobile:2008,Nobile:2008_2,sfem,gPC2002,gPC2003,xiu2009} and the references therein).

\section{Related Work and Background Review}

\subsection{Related Work on Density Estimation}
\label{related_work}
Let $x$ be a random variable, both kernel density estimation~\cite{Rosenblatt:1956,Parzen:1962,Hwang:1994} and asymptotic probability extraction~\cite{xli2004:Apex,Apex} aim to approximate its probability density function $\rho(x)$. 

{\it Kernel Density Estimation:} with $N$ samples for $x$, kernel density estimation approximates its probability density function by using a set of kernel functions. The probability density function generated by kernel density estimation is non-negative, and the resulting cumulative density function is bounded in $[0,1]$. However, kernel density estimation is seldom used in circuit modeling due to several shortcomings. First, the approximated probability density function is not compact: one has to store all samples as the parameters of a density function, which is inefficient for reuse in a stochastic simulator. Second, it is not straightforward to generate samples from the approximated probability density function. Third, the accuracy of kernel density estimation highly depends on the specific forms of the kernel functions (although Gaussian kernel seems suitable for the examples used in this work) as well as some parameters (e.g., the smoothing parameter). This paper will not construct the closed-form probability density function by kernel density estimation, instead we will use the \textit{numerical} results from kernel density estimation as a reference for accuracy comparison.

{\it Asymptotic Probability Extraction:} if $x$ is a linear quadratic function of some lower-level Gaussian parameters $\vec \xi$, asymptotic probability extraction~\cite{xli2004:Apex,Apex} can efficiently approximate $\rho(x)$ by moment matching. It has become the mainstream algorithm used in statistical circuit yield analysis and optimization. Asymptotic probability extraction and its variant~\cite{PEM} treat ${\rho} (x)$ as the impulse response of a linear time-invariant system, then approximates ${\rho} (x)$ using asymptotic waveform evaluation~\cite{AWE}. Several shortcomings have limited the application of asymptotic probability extraction and its variants:

1) Some assumptions of asymptotic probability extraction may not hold: i) $\vec \xi$ are assumed Gaussian variables, whereas in reality $\vec \xi$ can be non-Gaussian; ii) $x$ may not be a linear quadratic function of $\vec \xi$; iii) the statistical moments of $x$ may be unbounded, and thus asymptotic waveform evaluation~\cite{AWE} cannot be applied. 
 
2) The density functions from moment matching may be physically inconsistent. The cumulative density function may have oscillations and the probability density function may be negative, as shown by~\cite{xli2004:Apex,Apex} and the recent work~\cite{Krishnan:2013}, as well as by our experiments in Section~\ref{subsec:apexResults}. This is because the impulse response of a linear system is not guaranteed non-negative when generated by asymptotic waveform evaluation~\cite{AWE}. Negative probability density functions cannot be used for the stochastic simulation of a physical model.
 
3) The resulting density function may blow up, as shown in Section~\ref{subsec:apexResults} and in~\cite{Krishnan:2013}. There are two reasons for that. First, inaccurate moment computation can cause positive poles for a linear system, leading to an unbounded time-domain response. Second, asymptotic waveform evaluation is numerically unstable, which is well known in interconnect macromodeling. This is one of the important reasons why the model order reduction community has switched to implicit moment matching by Krylov-subspace projection.

\subsection{Generalized Polynomial-Chaos Basis Function and Gauss Quadrature}
\label{sec:back:gPC}
In order to apply stochastic spectral methods, one normally needs a set of generalized polynomial-chaos basis functions to approximate the stochastic solution of a physical model. Very often, a proper quadrature rule such as Gauss quadrature method~\cite{Golub:1969} is also required to set up a deterministic equation in intrusive solvers such as stochastic Galerkin~\cite{Pulch:2011,Pulch:2011_1} and stochastic testing~\cite{zzhang:tcad2013,zzhang:tcas2_2013,zzhang:iccad_2013}, or to recover the coefficients of each basis function for the solution in non-intrusive (i.e., sampling-based) solvers such as stochastic collocation~\cite{col:2005,Ivo:2007,Nobile:2008,Nobile:2008_2}.

{\it Basis Function Construction.} Given $\rho (x)$ (the probability density function of $x$), the generalized polynomial-chaos basis functions of $x$ are a set of orthonormal polynomials
	\begin{equation}
\label{uni_gPC}
 \int\limits_{\mathbb{R}}  {\phi_{i} ( {x } )\phi_{j} ( {x } ){\rho}( {x } )d x }=\delta_{i,j} 
\end{equation}
where integers $i$ and $j$ denote the polynomial degrees, and $\delta_{i,j}$ is a Delta function. In order to obtain $\phi_i(x)$'s, a set of orthogonal polynomials $\pi_i(x)$'s are first constructed via the well-known three-term recurrence relation~\cite{Walter:1982}:
\begin{equation}
\label{recurrence}
\begin{array}{l}
 \pi _{i + 1} (x) = \left( {x - \gamma _i } \right)\pi _i (x) - \kappa _i \pi _{i - 1} (x),\;\;i = 0,\;1, \cdots  \\ 
 \pi _{- 1} (x) = 0,\;\;\pi _0 (x) = 1,\;\; \\ 
 \end{array}
\end{equation}
where 
\begin{equation}
\label{int_cal}
\begin{array}{l}
 \gamma _i  = \frac{{\int\limits_{\mathbb{R}} {x\pi _i^2 (x)\rho (x)dx} }}{{\int\limits_{\mathbb{R}} {\pi _i^2 (x)\rho (x)dx} }}, \;\kappa _{i+1}  = \frac{{\int\limits_{\mathbb{R}} {\pi _{i+1}^2 (x)\rho (x)dx} }}{{\int\limits_{\mathbb{R}} {\pi _{i}^2 (x)\rho (x)dx} }},\;\;i = 0,1, \cdots, 
 \end{array}
\end{equation}
and $\kappa_0=1$. Here $\pi_i(x)$ is a degree-$i$ polynomial with leading coefficient 1. After that, the first $\hat {n}+1$ basis functions are obtained by normalization:
\begin{equation}
\phi _i (x) = \frac{{\pi _i (x)}}{{\sqrt {\kappa _0 \kappa _1  \cdots \kappa _i } }}, \; {\rm for}\; i=0,1,\cdots,\hat {n}.
\end{equation}
The obtained univariate basis functions can be easily extended to the multivariate cases, as detailed in Section II-A of~\cite{zzhang:iccad_2013}.

{\it Gauss-Quadrature Rule.} When computing an integral with Gauss quadrature~\cite{Golub:1969} one typically uses the expression
\begin{equation}
\label{stoInt}
\int\limits_{\mathbb{R} } {g( { x } ) \rho ( { x  } )dx }  \approx \sum\limits_{j = 1}^{\hat n+1} {g( { x^j } )} w^j
\end{equation}
which provides an exact result if $g\left( { x} \right)$ is a polynomial of degree $\leq 2\hat n+1$. The quadrature points ${ x^j }$'s and weights $w^j$'s depend on $\rho \left( {x } \right)$. Define a symmetric tridiagonal matrix
\begin{equation}
\textbf{J} = \left[ {\begin{array}{*{20}c}
   {\gamma _0 } & {\sqrt {\kappa _1 } } & {} & {} & {}  \\
   {\sqrt {\kappa _1 } } & {\gamma _1 } &  \ddots  & {} & {}  \\
   {} &  \ddots  &  \ddots  &  \ddots  & {}  \\
   {} & {} &  \ddots  & {\gamma _{\hat n - 1} } & {\sqrt {\kappa _{\hat n} } }  \\
   {} & {} & {} & {\sqrt {\kappa _{\hat n} } } & {\gamma _{\hat n} }  
\end{array}} \right],
\end{equation}
and let its eigenvalue decomposition be $\textbf{J} = \textbf{U}\Sigma \textbf{U}^T$, where $\textbf{U}$ is a unitary matrix. Denote the $(i,j)$ entry of $\textbf{U}$ by $u_{i,j}$, then $x^j$ is the $j$-th diagonal element of $\Sigma$, and the corresponding weight $w^j$ is $u_{1,j}^2$~\cite{Golub:1969}. Using tensor product or sparse grids, $1$-D Gauss quadrature rules can be easily extended to multi-dimensional cases~\cite{col:2005,Ivo:2007,Nobile:2008,Nobile:2008_2}.

{\it Remark 2.1:} The main bottleneck of the above procedures lies in (\ref{int_cal}), which requires computing a set of integrals. This step can be non-trivial if $\rho(x)$ is not in a proper form. When such integrals are not accurately computed, the constructed basis functions can be erroneous. Furthermore, $\kappa_i$'s may become negative, making computing the Gauss-quadrature points and weights impossible.

\section{The Proposed Framework}
\label{subsec:flow}
In this paper, $\hat x_i$'s in Fig.~\ref{fig:hierarchical} are assumed mutually independent. With this assumption, we can consider each surrogate model independently. For simplicity, let 
\begin{equation}
\label{surrogate}
\hat x=f(\vec \xi), \; {\rm with} \;\vec \xi\in \mathbb{R}^d
\end{equation}
represent a general surrogate model. We employ the linear transformation
\begin{equation}
\label{linear_trans}
x=\frac{\hat x-a}{b}
\end{equation}
to define a new random input $x$, which aims to improve the numerical stability. Once we obtain the cumulative density function and probability density function of $x$ (denoted as $p(x)$ and $\rho (x)$, respectively), then the cumulative density function and probability density function of $\hat x$ can be obtained by
\begin{equation}
\label{original_density}
\hat p (\hat x)=p\left(\frac{\hat x-a}{b}\right)\; {\rm and}\; \hat \rho (\hat x)=\frac{1}{b}\rho (\frac{\hat x-a}{b})
\end{equation}
respectively.
\begin{figure}[t]
	\centering
		\includegraphics[width=3.3in,height=45mm]{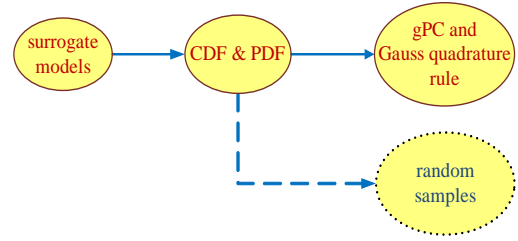} 
\caption{Construct generalized polynomial-chaos (gPC) bases and Gauss quadrature rules from surrogate models. Here CDF and PDF means ``cumulative density function" and ``probability density function", respectively.}
	\label{fig:gpc_flow}
\end{figure}

As shown in Fig.~\ref{fig:gpc_flow}, we first construct the density functions of $x$ {\it in a proper way}, then we determine the generalized polynomial-chaos bases of $x$ and a proper Gauss quadrature rule based on the obtained density functions. With the obtained cumulative density function, random samples of $x$ could be easily obtained for higher-level Monte Carlo-based simulation, however such task is not the focus of this paper. Our proposed framework consists of the following steps.
\begin{itemize}
	\item \textbf{Step 1.} Use $N$ Monte Carlo samples (or readily available measurement/simulation data) to obtain the discrete cumulative density function curve of $\hat x=f(\vec \xi)$. Since $f(\vec \xi)$ is a surrogate model, this step can be extremely efficient.
   	\item \textbf{Step 2.} Let $\delta>0$ be a small threshold value, $\hat x_{\rm min}$ and $\hat x_{\rm max}$ be the minimum and maximum values of $\hat x$ from the Monte Carlo analysis (or available data), respectively. We set $a$$=$$\hat x_{\rm min}-\delta$, $b$$=$$\hat x_{\rm max}+\delta-a$, then $N$ samples of $x$ in the interval $(0,1)$ are obtained by the linear transformation (\ref{linear_trans}). The obtained samples provide a discrete cumulative density function for $x$.
	\item \textbf{Step 3.} From the obtained cumulative density function curve of $x$, pick $n\ll N$ points $(x_i, y_i)$ for $i=1,\cdots,n$. Here $x_i$ denotes the value of $x$, and $y_i$ the corresponding cumulative density function value. The data are monotone: $x_i<x_{i+1}$ and $0=y_1\leq  \cdots \leq y_n=1$.
	\item \textbf{Step 4.} Use a monotone interpolation algorithm in Section~\ref{sec:fitting} to construct a closed-form function $p(x)$ to approximate the cumulative density function of $x$. 
	\item \textbf{Step 5.} Compute the first-order derivative of $p(x)$ and use it as a closed-form approximation to ${\rho}(x)$.
	\item \textbf{Step 6.} With the obtained ${\rho}(x)$, we utilize the procedures in Section~\ref{sec:mom} to construct the generalized polynomial-chaos basis functions and Gauss quadrature points/weights for $x$.
\end{itemize}

Many surrogate models are described by truncated generalized polynomial chaos expansions. The cost of evaluating such models may increase dramatically when the lower-level parameters $\vec \xi$ have a high dimensionality (which may occasionally happen), although the surrogate model evaluation is still much faster than the detailed simulation. Fortunately, in practical high-dimensional stochastic problems, normally only a small number of parameters are important to the output and most cross terms will vanish~\cite{xli2010,yxiu:2013,lyan:2012}. Consequently, a highly sparse generalized polynomial chaos expansion can be utilized for fast evaluation. Furthermore, when the coupling between the random parameters are weak, quasi-Monte Carlo~\cite{Sloan:quasiMC} can further speed up the surrogate model evaluation.

In Step 3, we first select $(x_1, y_1)=(0,0)$ and $(x_n, y_n)=(1,1)$. The $n$ data points are selected such that
\begin{equation}
\label{data:const}
\left| {x_{i + 1}  - x_i } \right| \le \frac{1}{m}\; {\rm and}\;\left| {y_{i + 1}  - y_i } \right| \le \frac{1}{m},
\end{equation}
where $m$ is an integer used to control $n$. This constraint ensures that the interpolation points are selected properly such that the behavior around the peak of $\rho (x)$ is well captured. In practical implementation, for $k=2, \cdots, n-1$, the point $(x_{k}, y_{k})$ is selected from the cumulative density function curve subject to the following criteria:
\begin{equation}
\label{data:arc}
\sqrt {(y_{k - 1}  - y_k )^2  + \left( {x_{k - 1}  - x_k } \right)^2} \approx \frac{1}{m}.
\end{equation}
For $x\notin [x_1,x_n]$, we set ${\rho}(x)$$=$$0$. This treatment introduces some errors in the tail regions. Approximating the tail regions is non-trivial, but such errors may be ignored if rare failure events are not a major concern (e.g., in the yield analysis of some analog/RF circuits).

{\it Remark 3.1:} Similar to standard stochastic spectral simulators~\cite{zzhang:tcad2013,zzhang:tcas2_2013,zzhang:iccad_2013,Pulch:2011,Pulch:2011_1,gPC2002,gPC2003,xiu2009,col:2005,Ivo:2007,Nobile:2008,Nobile:2008_2,sfem}, this paper assumes that $\hat x_i$'s are mutually independent. It is more difficult to handle correlated and non-Gaussian random inputs. Not only is it difficult to construct the density functions, but also it is hard to construct the basis functions even if the multivariate density function is given~\cite{UQ:book,arb_chaos}. How to handle correlated non-Gaussian random inputs remains an open and important topic in uncertainty quantification~\cite{UQ:book}. Some of our progress in this direction will be reported in~\cite{zzhang:huq}.

The most important parts of our algorithm are Step 4 and Step 6. In Section~\ref{sec:fitting} we will show how we guarantee that the obtained density functions are physically consistent. Step 6 will be detailed in Section~\ref{sec:mom}, with emphasis on an efficient analytical implementation.

\section{Implementation of the Density Estimator}
\label{sec:fitting}
This section presents the numerical implementation of our proposed density estimation. Our implementation is based on two monotone interpolation techniques, which are well studied in the mathematical community but have not been applied to uncertainty quantification. Since we approximate the cumulative density function $p(x)$ in the interval $x\in [x_1,x_n]$, in both methods we set $p(x)$$=$$y_1$$=$$0$ for $x<x_1$ and $p(x)$$=$$y_n$$=$$1$ for $x>x_n$, respectively.

\subsection{Method 1: Piecewise Cubic Interpolation}

Our first implementation uses a piecewise cubic interpolation~\cite{Fritsch:1980,Hyman:1983}. With the monotone data from Step 3 of Section~\ref{subsec:flow}, we construct $p(x)$ as a cubic polynomial:
\begin{equation}
\label{Px}
 p(x) = c_k^1  + c_k^2 (x - x_k ) + c_k^3 (x - x_k )^2  + c_k^4 (x - x_k )^3 
\end{equation}
for $x \in [x_k ,x_{k + 1} ]$, $0<k<n$. If $y_k$$=$$y_{k+1}$, we simply set $c_k^1$$=$$y_k$ and $c_k^2$$=$$c_k^3$$=$$c_k^4$$=0$. Otherwise, the coefficients are selected according to the following formula~\cite{Hyman:1983}
\begin{equation}
\begin{array}{l}
\label{coef:mpc}
 c_k^1  = y_k ,\;\;c_k^2  = \dot y_k ,\; \\ 
 c_k^3  = \frac{{s_k  - \dot y_{k + 1}  - 2\dot y_k }}{{\Delta x_k }},\;c_k^4  = \frac{{2s_k  - \dot y_{k + 1}  - \dot y_k }}{{\left( {\Delta x_k } \right)^2 }} \\ 
 \end{array}
\end{equation}
where $\Delta x_k$$=$$x_{k+1}$$-$$x_k$, $s_k$$=$$\frac{y_{k+1}-y_k}{\Delta x_k}$. This formula ensures that $p(x)$ and $p'(x)$ are continuous, $p(x_k)=y_k$ and $p'(x_k)=\dot y_k$. Here $p'(x)$ denotes the 1st-order derivative of $p(x)$.

The key of this implementation is how to compute $\dot y_k$ such that the interpolation is accurate and $p(x)$ is non-decreasing. The value of $\dot y_k $ is decided by two steps. First, we compute the first-order derivative $\dot y (x_k)$ by a parabolic method:
\begin{equation}
\label{para}
\dot y(x_k ) = \left\{ \begin{array}{l}
 \frac{{s_1 \left( {2\Delta x_1  + \Delta x_2 } \right) - s_2 \Delta x_1 }}{{x_3  - x_1 }},\;\;\;\;\;\;\;\;\;\;\;\;\;\;{\rm{if}}\;k = 1 \\ 
 \frac{{s_{n - 1} \left( {2\Delta x_{n - 1}  + \Delta x_{n - 2} } \right) - s_{n - 2} \Delta x_{n - 1} }}{{x_n  - x_{n - 2} }},\;\;{\rm{if}}\;k = n \\ 
 \frac{{s_k \Delta x_{k - 1}  + s_{k - 1} \Delta x_k }}{{x_{k + 1}  - x_{k - 1} }},\;\;\;\;\;\;\;\;\;\;\;{\rm{if}}\;2 < k < n - 1. \\ 
 \end{array} \right.
\end{equation}
This parabolic method has a $2$nd-order accuracy~\cite{Hyman:1983}. Second, $\dot y_k$ is obtained by perturbing $\dot y (x_k)$ (if necessary) to enforce the monotonicity of $p(x)$. The monotonicity of $p(x)$ is equivalent to $p'(x)\geq 0$, which is a $2$nd-order inequality. By solving this inequality, a feasible region for $\dot y_k$, denoted by ${\cal A}$, is provided in~\cite{Fritsch:1980}. Occasionally we need to project $\dot y (x_k)$ onto ${\cal A}$ to get $\dot y_k$ if $\dot y (x_k) \notin {\cal A}$. In practice, we use the simpler projection method suggested by~\cite{Hyman:1983}:
\begin{equation}
\label{proj}
\dot y_k  = \left\{ \begin{array}{l}
 \min \left( {\max \left( {0,\dot y(x_k )} \right),3s_{\min }^k } \right),\;{\rm{if }}\; s_k s_{k - 1}  > 0 \\ 
 0,\;\;{\rm{if }}\; s_k s_{k - 1}  = 0  
 \end{array} \right.
\end{equation}
with $s_0$$=$$s_{1}$, $s_n$$=$$s_{n-1}$ and $s_{\min }^k$$=$${\rm min} (s_k, s_{k-1})$. The above procedure projects $\dot y(x_k )$ onto a subset of ${\cal A}$, and thus the monotonicity of $p(x)$ is guaranteed.

Once $p(x)$ is constructed, the probability density function of $x$ can be obtained by
\begin{equation}
\label{pdfx}
\rho (x)=p'(x) = c_k^2 + 2c_k^3 (x - x_k )  + 3c_k^4 (x - x_k )^2
\end{equation}
for $x_k$$\leq x$$\leq x_{k+1}$. Note that for $x\notin [x_1,x_n]$, $p'(x)=0$. 

Calculating $p'(x)$ may amplify the interpolation errors. However, the error is acceptable since the constructed $p(x)$ is smooth enough and $p'(x)$ is continuous. The pseudo codes of Algorithm~\ref{alg:mpc} summarize the steps of this approach.
\begin{algorithm}[t]
\caption{piecewise cubic density estimation}
\label{alg:mpc}
\begin{algorithmic}[1]
\STATE {Evaluate the model~(\ref{surrogate}) to obtain $N$ samples of $\hat x$;}
\STATE {Shift and scale $\hat x$ to obtain $N$ samples for $x$;}
\STATE {Pick $n$ data points $(x_k,y_k)$, under constraint (\ref{data:const});} 
\STATE {Calculate ${\dot y (x_k)}$ using the parabolic method (\ref{para}); } 
\STATE {\textbf{for} $k=1,\cdots,  n$ \textbf{do} } 
\STATE {\hspace{5pt}\textbf{if} $y_k=y_{k+1}$, set $c_k^1$$=$$y_k$ and $c_k^2$$=$$c_k^3$$=$$c_k^4$$=0$; } 
\STATE {\hspace{5pt}\textbf{else}  } 
\STATE {\hspace{10pt}Compute $\dot y_k$ according to (\ref{proj});} 
\STATE {\hspace{10pt}Compute the coefficients in (\ref{coef:mpc}).}
\STATE {\hspace{5pt}\textbf{end} }
\STATE {\textbf{end for}} 
\end{algorithmic}
\end{algorithm}

\subsection{Method 2: Piecewise Rational Quadratic Interpolation}
Our second implementation is based on a piecewise rational quadratic interpolation~\cite{Gregory:1985,Gregory:1982}. In this implementation, we approximate the cumulative density function of $x$ by 
\begin{equation}
\label{mprq}
p(x) = \frac{{N(x)}}{{D(x)}} = \frac{{\alpha _k^1  + \alpha _k^2 x + \alpha _k^3 x^2 }}{{\beta _k^1  + \beta _k^2 x + \beta _k^3 x^2 }}
\end{equation}
for $x \in [x_k ,x_{k + 1} ]$. The coefficients are selected by the following method: when $x_k=x_{k+1}$, we set $\alpha _k^1=y_k$, $\beta _k^1=1$ and all other coefficients to zero; otherwise, the coefficients are decided according to the formula
\begin{equation}
\label{cofmprq}
\begin{array}{l}
 \alpha _k^1  = y_{k + 1} x_k^2  - w_k x_k x_{k + 1}  + y_k x_{k + 1}^2 , \\ 
 \alpha _k^2  = w_k (x_k  + x_{k + 1} ) - 2y_{k + 1} x_k  - 2y_k x_{k + 1} , \\
  \alpha _k^3  = y_{k + 1}  - w_k  + y_k ,\; \\ 
 \beta _k^1  = x_k^2  - v_k x_k x_{k + 1}  + x_{k + 1}^2 , \\ 
 \beta _k^2  = v_k (x_k  + x_{k + 1} ) - 2x_k  - 2x_{k + 1} ,  \; \beta _k^3  = 2 - v_k ,\; \\ 
 {\rm{with}}\;w_k  = \frac{{y_{k + 1} \dot y_k  + y_k \dot y_{k + 1} }}{{s_k }}\;{\rm and}\; v_k  = \frac{{\dot y_k  + \dot y_{k + 1} }}{{s_k }} 
 \end{array}
\end{equation}
where $s_k$ is defined the same as in piecewise cubic interpolation. In this interpolation scheme, the sufficient and necessary condition for the monotonicity of $p(x)$ is very simple: $\dot y_k\geq 0$. In order to satisfy this requirement, the slope $\dot y_k$ is approximated by the geometric mean
\begin{equation}
\label{geomean}
\dot y_k  = \left\{ \begin{array}{l}
 \left( {s_1 } \right)^{\frac{{x_3  - x_1 }}{{x_3  - x_2 }}} \left( {s_{3,1} } \right)^{\frac{{x_1  - x_2 }}{{x_3  - x_2 }}} ,\;\;{\rm{if}}\;k = 1 \\ 
 \left( {s_{n - 1} } \right)^{\frac{{x_n  - x_{n - 2} }}{{x_{n - 1}  - x_{n - 2} }}} \left( {s_{n,n - 2} } \right)^{\frac{{x_{n - 1}  - x_n }}{{x_{n - 1}  - x_{n - 2} }}} ,\;\;{\rm{if}}\;k = n \\ 
 \left( {s_{k - 1} } \right)^{\frac{{x_{k + 1}  - x_k }}{{x_{k + 1}  - x_{k - 1} }}} \left( {s_k } \right)^{\frac{{x_k  - x_{k - 1} }}{{x_{k + 1}  - x_{k - 1} }}} ,\;{\rm{if}}\; 1< k <n 
 \end{array} \right.
\end{equation}
with $s_{k_1,k_2}=\frac{y_{k_1}-y_{k_2}}{x_{k_1}-x_{k_2}}$. Similarly, the probability density function of $x$ can be approximated by
\begin{equation}
\label{mprq:pdf}
\rho (x)=p'(x) = \frac{{N'(x)D(x)-D'(x)N(x)}}{{D^2(x)}},
\end{equation}
for $x \in [x_k ,x_{k + 1} ]$. 

Note that in piecewise cubic interpolation, a projection procedure is not required, since the monotonicity of $p(x)$ is automatically guaranteed. The pseudo codes of this density estimation method are provided in Algorithm~\ref{alg:mprq}.
\begin{algorithm}[t]
\caption{piecewise rational quadratic density estimation}
\label{alg:mprq}
\begin{algorithmic}[1]
\STATE {Evaluate the model~(\ref{surrogate}) to obtain $N$ samples of $x$;}
\STATE {Shift and scale $\hat x$ to obtain $N$ samples for $x$;}
\STATE {Pick $n$ data points $(x_k,y_k)$, under constraint (\ref{data:const});} 
\STATE {\textbf{for} $k=1,\cdots,  n$ \textbf{do} } 
\STATE {\hspace{5pt}Calculate ${\dot y_k}$ using the formula in (\ref{geomean}); } 
\STATE {\hspace{5pt}\textbf{if} $y_k=y_{k+1}$ } 
\STATE {\hspace{5pt} set $\alpha _k^1=y_k$, $\beta _k^1=1$ and other coefficients to zero; } 
\STATE {\hspace{5pt}\textbf{else} } 
\STATE {\hspace{10pt} compute the coefficients of $N(x)$ and $D(x)$ using (\ref{cofmprq}). } 
\STATE {\hspace{5pt}\textbf{end} }
\STATE {\textbf{end for}} 
\end{algorithmic}
\end{algorithm}

\subsection{Properties of $p(x)$}
\label{subsec:px_property}
It is straightforward to show that the obtained density functions are physically consistent: 1) $p(x)$ is differentiable, and thus its derivative $p'(x)$ always exists; 2) $p(x)$ is monotonically increasing from $0$ to $1$, and the probability density function $\rho(x)$ is non-negative.

We can easily draw a random sample from the obtained $p(x)$. Let $y\in [0,1]$ be a sample from a uniform distribution, then a sample of $x$ can be obtained by solving $p(x)=y$ in the interval $y\in [y_k, y_{k+1}]$. This procedure only requires computing the roots of a cubic (or quadratic) polynomials, resulting in a unique solution $x\in [x_k, x_{k+1}]$. This property is very useful in uncertainty quantification. Not only are random samples used in Monte Carlo simulators, but also they can be used in stochastic spectral methods. Recently, compressed sensing has been applied to high-dimensional stochastic problems~\cite{xli2010,yxiu:2013,lyan:2012}. In compressed sensing, random samples are normally used to enhance the restricted isometry property of the dictionary matrix~\cite{Donoho:2006}.

Finally, it becomes easy to determine the generalized polynomial-chaos basis functions and a proper quadrature rule for $x$ due to the special form of $\rho(x)$. This issue will be discussed in Section~\ref{sec:mom}.

{\it Remark 4.1:} Our proposed density estimator only requires some interpolation points from a discrete cumulative density function curve. The interpolation points actually can be obtained by any appropriate approach. For example, kernel density estimation will be a good choice if we know a proper kernel function and a good smoothing parameter based on {\it a-priori} knowledge. When the surrogate model is a linear quadratic function of Gaussian variables, we may first employ asymptotic probability extraction~\cite{xli2004:Apex} to generate a physically inconsistent cumulative density function. After that, some monotone data points (with $y_i$'s bounded by $0$ and $1$) can be selected to generate a piecewise cubic or piecewise rational quadratic cumulative density function. The new cumulative density function and probability density function become physically consistent and can be reused in a stochastic simulator.

\section{Determine Basis Functions and Gauss Quadrature Rules}
\label{sec:mom}

This section shows how to calculate the generalized polynomial-chaos bases and the Gauss quadrature points/weights of $x$ based on the obtained density function.

\subsection{Proposed Implementation}
One of the many usages of our density estimator is to fast compute a set of generalized polynomial-chaos basis functions and Gauss quadrature points/weights by analytically computing the integrals in (\ref{int_cal}). Let $\pi _i^2 (x) = \sum\limits_{k = 0}^{2i} {\tau _{i,k} x^k }$, then we have
\begin{equation} 
\begin{array}{l}
 \int\limits_{\mathbb{R}} {x\pi _i^2 (x)\rho (x)dx}  = \sum\limits_{k = 0}^{2i} {\tau _{i,k} M_{k + 1} } , \\ 
 \int\limits_{\mathbb{R}}  {\pi _i^2 (x)\rho (x)dx}  = \sum\limits_{k = 0}^{2i} {\tau _{i,k} M_k }  
 \end{array}
\end{equation}
where $M_k$ denotes the $k$-th statistical moments of $x$. By exploiting the special form of our obtained density function, the statistical moments can be computed as
\begin{equation}
M_k  = \int\limits_{-\infty }^{+\infty } {x^k \rho(x)dx}=\int\limits_{x_1 }^{x_n } {x^k \rho(x)dx}  = \sum\limits_{j = 1}^{n - 1} {I_{j,k} } 
\end{equation}
where $I_{j,k}$ denotes the integral in the $j$-th piece:
\begin{equation}
I_{j,k}  = \int\limits_{x_j }^{x_{j + 1} } {x^k \rho(x)dx} =F_{j,k}(x_{j+1})-F_{j,k}(x_j).
\end{equation}
Here $F_{j,k}(x)$ is a continuous analytical function under the constraint $\frac{d}{dt}F_{j,k}(x)=x^k\rho(x)$ for $x\in [x_j,x_{j+1}]$. 
The key problem of our method is to construct $F_{j,k}(x)$.  When $\rho(x)$ is obtained from Alg. 1 or Alg. 2, we can easily obtain the closed form of $F_{j,k} (x)$, as will be elaborated in Section~\ref{mom:alg1} and Section~\ref{mom:alg2}.

{\it Remark 5.1:} This paper directly applies (\ref{int_cal}) to compute the recurrence parameters $\gamma_i$ and $\kappa_i$. As suggested by~\cite{Walter:1982}, modified Chebyshev algorithm~\cite{Wheeler:1974} can improve the numerical stability when constructing high-order polynomials. Modified Chebyshev algorithm indirectly computes $\gamma_i$ and $\kappa_i$ by first evaluating a set of modified moments. Again, if we employ the $\rho(x)$ obtained from our proposed density estimators, then the calculation of modified moments can also be done analytically to further improve the accuracy and numerical stability.

\subsection{Construct $F_{j,k} (x)$ using the Density Function from Alg. 1}
\label{mom:alg1}
When $\rho(x)$ is constructed by Alg. 1, $x^k\rho(x)$ is a polynomial function of at most degree $k+2$ inside the interval $[x_j, x_{j+1}]$. Therefore, the analytical form of $F_{j,k}(x)$ is
\begin{equation}
F_{j,k} (x) =\textbf{a}_{j,k} x^{k + 3}  + \textbf{b}_{j,k} x^{k + 2}  + \textbf{c}_{j,k} x^{k + 1} 
\end{equation}
with 
\begin{equation}
\begin{array}{l}
 \textbf{a}_{j,k}  = \frac{{3c_j^4 }}{{k + 3}},\;\;\textbf{b}_{j,k}  = \frac{{2c_j^3  - 6c_j^4 x_j }}{{k + 2}},\; \\ 
 \textbf{c}_{j,k}  = \frac{{c_j^2  - 2c_j^3 x_j  + 3c_j^4 x_j^2 }}{{k + 1}}. \\ 
 \end{array}\nonumber
\end{equation}

\subsection{Construct $F_{j,k} (x)$ using the Density Function from Alg. 2}
\label{mom:alg2}
If $\rho(x)$ is constructed by Alg. 2, for any $x\in [x_j, x_{j+1}]$ we rewrite $x^k\rho(x)$ as follows
\begin{equation}
\begin{array}{l}
 x^k \rho(x) = \frac{{x^k \left[ {N'(x)D(x) - D'(x)N(x)} \right]}}{{D^2 (x)}} \\ 
 \;\;\;\;\;\;\;\;\;\;\;\; = \frac{d}{{dx}}\left( {\frac{{x^k N(x)}}{{D(x)}}} \right) - \frac{{kx^{k - 1} N(x)}}{{D(x)}}. \\ 
 \end{array} \nonumber
\end{equation}
Therefore, $F_{j,k}(x)$ can be selected as
\begin{equation}
F_{j,k} (x) = \frac{{x^k N(x)}}{{D(x)}} - \tilde F_{j,k} (x),\;{\rm{with}}\;\frac{d}{{dx}}\tilde F_{j,k} (x) = \frac{{kx^{k - 1} N(x)}}{{D(x)}}.\nonumber
\end{equation}

In order to obtain $\tilde F_{j,k}(x)$, we perform a long division:
\begin{equation}
\frac{{kx^{k - 1} N(x)}}{{D(x)}} = \tilde P_{j,k} (x) + \frac{{\tilde R_{j,k} (x)}}{{D(x)}}
\end{equation}
where $\tilde P_{j,k} (x)$ and $\tilde R_{j,k} (x)$ are both polynomial functions, and $\tilde R_{j,k} (x)$ has a lower degree than $D(x)$. Consequently, 
\begin{equation}
\tilde F_{j,k} (x) = \tilde F_{j,k}^1 (x) + \tilde F_{j,k}^2 (x)
\end{equation}
where $\tilde F_{j,k}^1 (x)$ and $\tilde F_{j,k}^2 (x)$ are the integrals of $\tilde P_{j,k} (x)$ and $\frac{{\tilde R_{j,k} (x)}}{{D(x)}}$, respectively. It is trivial to obtain $\tilde F_{j,k}^1 (x)$ since $\tilde P_{j,k} (x)$ is a polynomial function. 
\begin{figure*}[!t]
\normalsize
\begin{equation}
\label{int:2ndRat}
\tilde F_{j,k}^2 (x) = \left\{ \begin{array}{l}
 \frac{{\tilde r_{j,k}^1 }}{{2\beta _j^3 }}\ln \left| {\beta _j^3 x^2  + \beta _j^2 x + \beta _j^1 } \right| + \frac{{2\beta _j^3 \tilde r_{j,k}^0  - \beta _j^2 \tilde r_{j,k}^1 }}{{\beta _j^3\sqrt {\Delta _j } }}\arctan \frac{{2\beta _j^3 x + \beta _j^2 }}{{\sqrt {\Delta _j } }},\;\;{\rm{if}}\;\Delta _j  > 0\; \\ 
 \frac{{\tilde r_{j,k}^1 }}{{2\beta _j^3 }}\ln \left| {\beta _j^3 x^2  + \beta _j^2 x + \beta _j^1 } \right| - \frac{{2\beta _j^3 \tilde r_{j,k}^0  - \beta _j^2 \tilde r_{j,k}^1 }}{{\beta _j^3\sqrt { - \Delta _j } }}\arctan \frac{{2\beta _j^3 x + \beta _j^2 }}{{\sqrt { - \Delta _j } }},\;\;{\rm{if}}\;\Delta _j  < 0 \\ 
 \frac{{\tilde r_{j,k}^1 }}{{2\beta _j^3 }}\ln \left| {\beta _j^3 x^2  + \beta _j^2 x + \beta _j^1 } \right| - \frac{{2\beta _j^3 \tilde r_{j,k}^0  - \beta _j^2 \tilde r_{j,k}^1 }}{{\beta _j^3\left( {2\beta _j^3 x + \beta _j^2 } \right)}},\;\;\;\;\;\;\;\;\;\;\;\;\;\;\;\;\;\;\;\;\;\;\;\;\;{\rm{if}}\;\Delta _j  = 0\; \\ 
 \end{array} \right.
\end{equation}
\setcounter{MYtempeqncnt}{\value{equation}}
\hrulefill
\vspace*{4pt}
\end{figure*}

The closed form of $\tilde F_{j,k}^2 (x)$ is decided according to the coefficients of $D(x)$ and $\tilde R_{j,k} (x)$, as is summarized below.

\textit{Case 1:} if $\beta_j^3\neq 0$, then $\tilde R_{j,k} (x)=\tilde r_{j,k}^0+\tilde r_{j,k}^1x$. Let us define $\Delta_j:=4\beta_j^1 \beta_j^3-\beta_j^2$, then we can select $\tilde F_{j,k}^2(x)$ according to the formula in (\ref{int:2ndRat}).

\textit{Case 2:} if $\beta_j^3= 0$ and $\beta_j^2\neq 0$, then $\tilde R_{j,k} (x)=\tilde r_{j,k}^0$ is a constant. In this case, we select
\begin{equation}
\tilde F_{j,k}^2 (x) = \frac{{\tilde r_{j,k}^0 }}{{\beta _j^2 }}\ln \left| {\beta _j^2 x + \beta _j^1 } \right|.
\end{equation}

\textit{Case 3:} if $\beta_j^3= \beta_j^2= 0$, then $\tilde R_{j,k} (x)=0$. In this case we set $\tilde F_{j,k}^2 (x)=0$.

{\it Remark 5.2:} Occasionally, the projection procedure (\ref{proj}) in Alg. 1 may cause extra errors at the end points of some intervals. If this problem happens we recommend to use Alg. 2. On the other hand, if high-order basis functions is required we recommend Alg. 1, since the moment computation with the density from Alg. 2 is numerically less stable (due to the long-term division and the operations in (\ref{int:2ndRat}).

\section{Numerical Examples}
This section presents the numerical results on a synthetic example and the statistical surrogate models from two practical analog/RF circuits. The surrogate models of these practical circuits are extracted from transistor-level simulation using the fast stochastic circuit simulator developed in~\cite{zzhang:tcad2013,zzhang:tcas2_2013,zzhang:iccad_2013}. All experiments are run in Matlab on a 2.4GHz 4-GB RAM laptop.

In the following experiments, we use the density functions from kernel density estimation as the ``reference solution" because: 1) as a standard technique, kernel density estimation is most widely used in mathematics and engineering; 2) kernel density estimation guarantees that the generated probability density function is non-negative, whereas asymptotic probability extraction cannot; 3) Gaussian kernel function seems to be a good choice for the examples in this paper. However, it is worth noting that the density functions from kernel density estimation are not efficient for reuse in higher-level stochastic simulation. We plot the density functions of $\hat x$ (the original random input) instead of $x$ (the new random input after a linear transformation) since the original one is physically more intuitive. To verify the accuracy of the computed generalized polynomial-chaos bases and Gauss quadrature points/weights, we define a symmetric matrix $\textbf{V}_{\hat n+1} \in \mathbb{R}^{(\hat n+1) \times (\hat n+1)}$, the $(i,j)$ entry of which is
\begin{equation}
\label{inner_matrix}
v_{i,j}  = \sum\limits_{k = 1}^{\hat n + 1} {w^k \phi _{i - 1} \left( {x^k } \right)\phi _{j - 1} \left( {x^k } \right)}. \nonumber
\end{equation}
Here $x^k$ and and $w^k$ are the computed $k$-th Gauss quadrature point and weight, respectively. Therefore $v_{i,j}$ approximates the inner product of $\phi _{i - 1}(x)$ and $\phi _{j - 1}(x)$, defined as $\int\limits_{\mathbb{R}}  {\phi _{i - 1} \left( {x } \right)\phi _{j - 1} \left( {x } \right)\rho (x)dx }$, by $\hat n+1$ quadrature points. Let $\textbf{I}_{\hat n+1}$ be an identity matrix, then we define an error:
\begin{equation}
\label{gpc_error}
\epsilon=||\textbf{I}_{\hat n+1}-\textbf{V}_{\hat n+1} ||_{\infty}
\end{equation}
which is close to zero when our constructed basis functions and Gauss-quadrature points/weights are accurate enough.

\subsection{Synthetic Example}
As a demonstration, we first consider the following synthetic example with four random parameters $\vec \xi=[\xi_1, \cdots, \xi_4]$:
\begin{equation}
\nonumber
\begin{array}{l}
 \hat x = f(\vec \xi ) = \xi _1  + 0.5\exp (0.52\xi _2 ) \\ 
 \;\;\;\;\;\;\;\;\;\;\;\;\;\;\;\;\;+ 0.3\sqrt {2.1 \times \left| {\xi _4 } \right|}  + \sin \left( {\xi _3 } \right)\cos \left( {3.91\xi _4 } \right) \\ 
 \end{array}
\end{equation}
where $\xi_1, \xi_2$ and $\xi_3$ are all standard Gaussian random variables, and $\xi_4$ has a uniform distribution in the interval $[-0.5, 0.5]$. This model is strongly nonlinear with respect to $\vec \xi$ due to the exponential, triangular and square root functions. It is also non-smooth at $\xi_4=0$ due to the third term in the model. This model is designed to challenge our algorithm. Using this surrogate model, $10^6$ samples of $x$ are easily created to generate the cumulative density function curve within $1$ second.
\begin{figure}[t]
	\centering
		\includegraphics[width=3.3in]{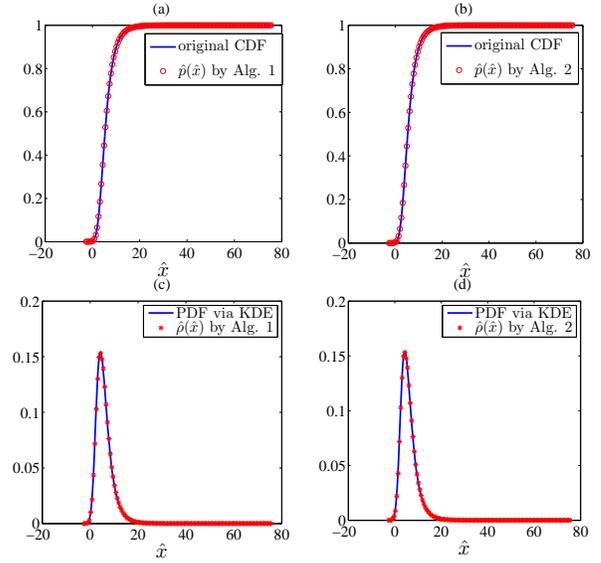} 
\caption{Cumulative density function (CDF) and probability density function (PDF) approximation of $\hat x$ for the synthetic example. The reference PDF is generated by kernel density estimation (KDE).}
	\label{fig:syn_density}
\end{figure}

\textit{Density Estimation:} we set $m=45$ and select $74$ data points from the obtained cumulative density function curve using the constraint in (\ref{data:arc}). After that, both Alg. 1 and Alg. 2 are applied to generate $p(x)$ and $\rho(x)$ as approximations to the cumulative density function and probability density function of $x$, respectively. The CPU times cost by our proposed density estimators are in millisecond scale, since only simple algebraic operations are required. After scaling by (\ref{original_density}), the cumulative density function and probability density function of the original random input $\hat x$ ($\hat p(\hat x)$ and $\hat \rho(\hat x)$, respectively) from both algorithms are compared with the original cumulative density function and probability density function in Fig.~\ref{fig:syn_density}. Clearly, $\hat p(\hat x)$ is indistinguishable with the original cumulative density function (from Monte Carlo simulation); and $\hat \rho(\hat x)$ overlaps with the original probability density function (estimated by kernel density estimation using Gaussian kernels). Note that the results from kernel density estimation are not efficient for reuse in higher-level stochastic simulation, since all Monte Carlo samples are used as parameters of the resulting density function.

It is clearly shown that the generated $\hat p(\hat x)$ [and thus $p(x)$] is monotonically increasing from $0$ to $1$, and that the generated $\hat \rho(\hat x)$ [and thus $\rho(x)$] is non-negative. Therefore, the obtained density functions are physically consistent.

\begin{figure}[t]
	\centering
		\includegraphics[width=3.3in]{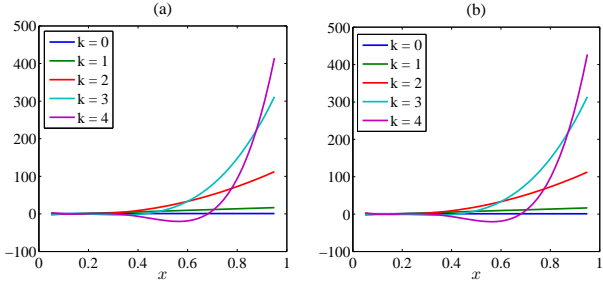} 
\caption{Computed generalized polynomial-chaos basis functions $\phi_k(x)$ ($k=0,\cdots,4$) for the synthetic example. (a) uses the probability density function from Alg. 1, and (b) uses the probability density function from Alg. 2.}
	\label{fig:syn_gpc}
\end{figure}

\textit{Basis Function:} Using the obtained density functions and the proposed implementation in Section~\ref{sec:mom}, a set of orthonormal polynomials $\phi_k(x)$'s are constructed as the basis functions at the cost of milliseconds. Fig.~\ref{fig:syn_gpc} show the first five generalized polynomial-chaos basis functions. Note that although the computed basis functions from two methods are graphically indistinguishable, they are actually slightly different since Alg. 1 and Alg. 2 generate different representations for $\rho(x)$.
\begin{table}[t]
	\centering
	\caption{Computed Gauss quadrature points and weights for the synthetic example.}	
	\label{tab:Gauss_syn}
	\smallskip 
  \begin{threeparttable}
		\begin{tabular}{|c|c|c|c|}
	\hline
		\multicolumn{2}{|c|}{with $\rho(x)$ from Alg. 1}	& \multicolumn{2}{|c|}{with $\rho(x)$ from Alg. 2} \\ \hline
		$x^k$ & $w^k$ & $x^k$ & $w^k$ \\ \thickhline				
		0.082620 &0.311811 & 0.084055 & 0.332478\\  \hline		
		 0.142565 &0.589727 & 0.144718 & 0.576328\\  \hline
		0.249409 &0.096115 & 0.252980 & 0.089027\\  \hline
		0.458799 &0.002333 & 0.463207 & 0.002150\\  \hline
		0.837187 &0.000016 & 0.835698 & 0.000016\\  \hline
	\end{tabular} 	
\end{threeparttable}	
\end{table}

\textit{Gauss Quadrature Rule:} setting $\hat n=4$, five Gauss quadrature points and weights are generated using the method presented in Section~\ref{sec:mom}. Table~\ref{tab:Gauss_syn} shows the results from two kinds of approximated density functions. Clearly, since the probability density functions from Alg. 1 and Alg. 2 are different, the resulting quadrature points/weights are also slightly different. The results from both probability density functions are very accurate. Using the probability density function from Alg. 1, we have $\epsilon=2.24\times 10^{-14}$, and the error (\ref{gpc_error}) is $7.57\times 10^{-15}$ if $\rho(x)$ from Alg. 2 is employed.

\subsection{Colpitts Oscillator}
\begin{figure}[t]
	\centering
		\includegraphics[width=3.3in]{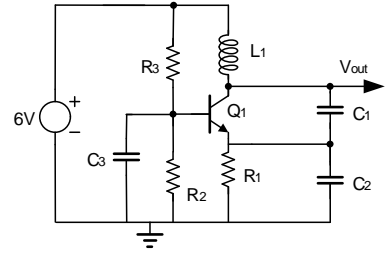} 
\caption{Schematic of the Colpitts oscillator.}
	\label{fig:Col_osc}
\end{figure}
\begin{figure}[t]
	\centering
		\includegraphics[width=3.3in]{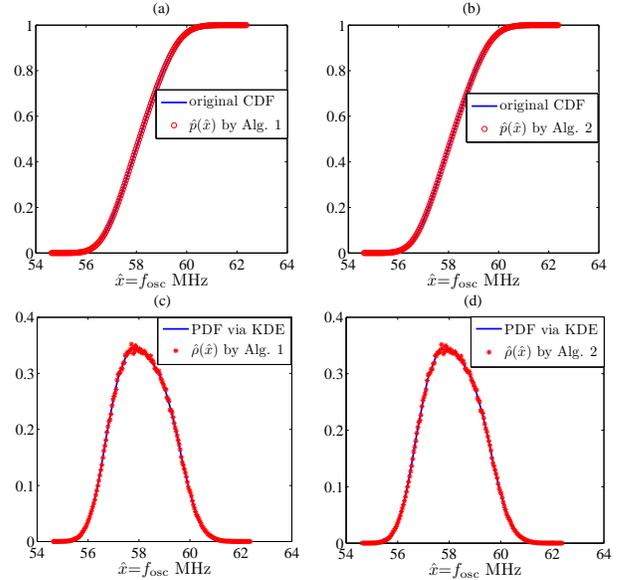} 
\caption{Cumulative density function (CDF) and probability density function (PDF) approximation for the frequency of the Colpitts oscillator. The reference PDF is generated by kernel density estimation (KDE).}
	\label{fig:osc_density}
\end{figure}	
We now test our proposed algorithm on a more practical example, the Colpitts oscillator circuit shown in Fig.~\ref{fig:Col_osc}. The design parameters of this circuit are $R_1$$=$$ 2.2$ k$\Omega$, $R_2$$=$$R_3$$=$$10$ k$\Omega$, $C_2$$=$$100$ pF, $C_3$$=$$0.1 \mu$F, and $\alpha$$=$$0.992$ for the BJT. The oscillation frequency is mainly determined by the values of $L_1$, $C_1$ and $C_2$. In this circuit, $L_1$$=$$150+{\cal N}(0,9)$ nH and $C_1$$=$$100+{\cal U}(-10,10)$ pF are random variables with Gaussian and uniform distributions, respectively. We construct a surrogate model using generalized polynomial chaos expansions and the stochastic shooting Newton solver in~\cite{zzhang:tcas2_2013}. The oscillation frequency $f_{\rm osc}$ is expressed as
\begin{equation}
\label{model:osc}
\hat {x}=f_{\rm osc} =f(\vec \xi) =\frac{1}{{\sum\limits_{k = 1}^{10} {T_k \psi _k ( {\vec \xi } )} }}
\end{equation}
where the denominator is a 3rd-order generalized polynomial chaos representation for the period of the oscillator, with $\psi _k ( {\vec \xi } )$ being the $k$-th multivariate generalized polynomial-chaos basis function of $\vec \xi$ and $T_k$ the corresponding coefficient. Although the period is a polynomial function of $\vec \xi$, the frequency is not, due to the inverse operation. In order to extract the cumulative density function curve, $5\times 10^5$ samples are utilized to evaluate the surrogate model (\ref{model:osc}) by Monte Carlo, which costs $225$ seconds of CPU times on our Matlab platform.

\textit{Density Estimation:} $106$ data points on the obtained cumulative density function curve are used to construct $p(x)$ and $\rho(x)$, which costs only several milliseconds. After scaling the constructed closed-form cumulative density functions and probability density functions from Alg. 1 and Alg. 2, the approximated density functions of the oscillation frequency are compared with the Monte Carlo results in Fig.~\ref{fig:osc_density}. The constructed cumulative density functions by both methods are graphically indistinguishable with the result from Monte Carlo. The bottom plots in Fig.~\ref{fig:osc_density} also show a good match between our obtained $\hat \rho(\hat x)$ with the result from kernel density estimation. Again, important properties of the density functions (i.e., monotonicity and boundedness of the cumulative density function, and non-negativeness of the probability density function) are well preserved by our proposed density estimation algorithms. 

\begin{figure}[t]
	\centering
		\includegraphics[width=3.3in]{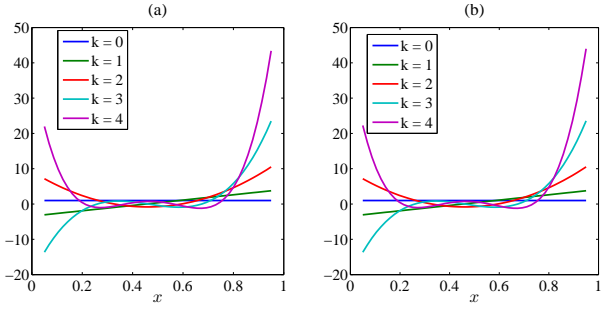} 
\caption{Computed generalized polynomial-chaos basis functions $\phi_k(x)$ ($k=0,\cdots,4$) for the Colpitts oscillator. (a) uses the probability density function from Alg. 1, and (b) uses the probability density function from Alg. 2.}
	\label{fig:osc_gpc}
\end{figure}

\textit{Basis Function:} Using the obtained density functions and the proposed implementation in Section~\ref{sec:mom}, a set of orthonormal polynomials $\phi_k(x)$'s are constructed as the basis functions at the cost of milliseconds. Fig.~\ref{fig:osc_gpc} shows several generalized polynomial-chaos basis functions of $x$. Again, the basis functions resulting from our two density estimation implementations are only slightly different.

\textit{Gauss Quadrature Rule:} the computed five Gauss quadrature points and weights are shown in Table~\ref{tab:Gauss_osc}. Again the results from two density estimations are slightly different. The results from both probability density functions are very accurate. Using $\rho(x)$ from Alg. 1, we have $\epsilon=1.3\times 10^{-13}$, and the error is $1.45\times 10^{-13}$ if we use $\rho(x)$ from Alg. 2.
\begin{table}[t]
	\centering
	\caption{Computed Gauss quadrature points and weights for the Colpitts oscillator.}
	\label{tab:Gauss_osc}
	\smallskip 
  \begin{threeparttable}
		\begin{tabular}{|c|c|c|c|}
	\hline
		\multicolumn{2}{|c|}{with $\rho(x)$ from Alg. 1}	& \multicolumn{2}{|c|}{with $\rho(x)$ from Alg. 2} \\ \hline
		$x^k$ & $w^k$ & $x^k$ & $w^k$ \\ \thickhline				
		0.170086 &0.032910 & 0.170935 & 0.032456\\  \hline		
		 0.309764 &0.293256 & 0.310016 & 0.292640\\  \hline
		0.469034 &0.441303 & 0.468658 & 0.439710\\  \hline
		0.632232 &0.217359 & 0.631249 & 0.218274\\  \hline
		0.788035 &0.016171 & 0.786226 & 0.016820\\  \hline
	\end{tabular} 	
\end{threeparttable}	
\end{table}

\subsection{Low-Noise Amplifier}
\begin{figure}[t]
	\centering
		\includegraphics[width=3.3in]{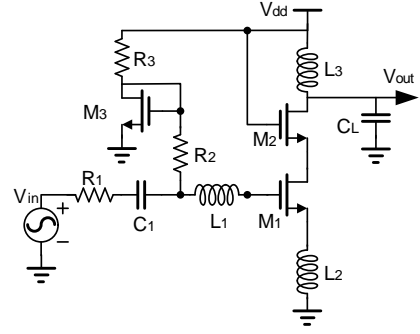} 
\caption{Schematic of the low-noise amplifier.}
	\label{fig:LNA}
\end{figure}
\begin{figure}[t]
	\centering
		\includegraphics[width=3.3in]{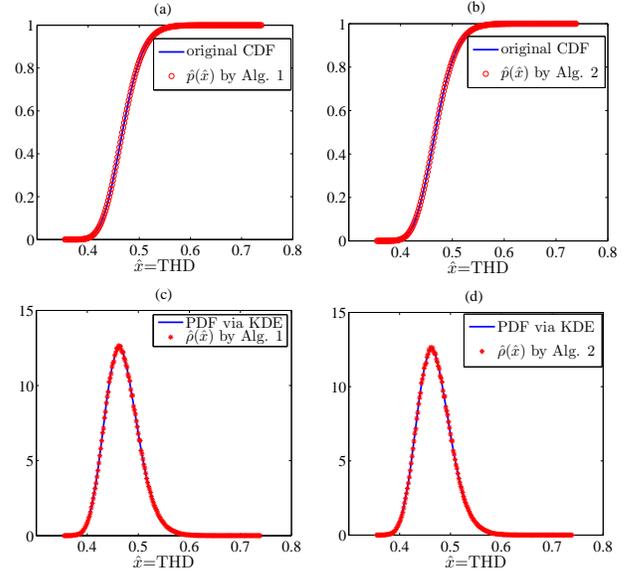} 
\caption{Cumulative density function (CDF) and probability density function (PDF) for the total harmonic distortion (THD) of the low-noise amplifier. The reference PDF is generated by kernel density estimation (KDE).}
	\label{fig:LNA_density}
\end{figure} 
In this example we consider the statistical behavior of the total harmonic distortion at the output node of the low-noise amplifier shown in Fig.~\ref{fig:LNA}. The device ratios of the MOSFETs are $W_1/L_1$$=$$W_2/L_2$$=$$500/0.35$ and $W_3/L_3$$=$$50/0.35$. The linear components are $R_1$$=$$50 \Omega$, $R_2$$=$$2$ k$\Omega$, $C_1$$=$$10$ pF, $C_L$$=$$0.5$ pF, $L_1$$=$$20$ nH and $L_3$$=$$7$ nH. Four random parameters are introduced to describe the uncertainties: $\xi_1$ and $\xi_2$ are standard Gaussian variables, $\xi_3$ and $\xi_4$ are standard uniform-distribution parameters. These random parameters are mapped to the physical parameters as follows: temperature $T$$=$$300+40\xi_1$ K influences transistor threshold voltage; $V_{\rm T}$$=$$0.4238+0.1\xi_2$ V represents the threshold voltage under zero $V_{\rm bs}$; $R_3$$=$$0.9+0.2\xi_3$ k$\Omega$ and $L_2$$=$$0.8+1.2\xi_4$ nH. The supply voltage is $V_{\rm dd}$$=$$1.5$ V, and the periodic input is $V_{\rm in}=0.1{\rm sin}(4\pi\times 10^8t)$ V. 

The surrogate model for total harmonic distortion analysis is constructed by a numerical scheme as follows. First, the parameter-dependent periodic steady-state solution at the output is solved by the non-Monte Carlo simulator in~\cite{zzhang:tcas2_2013}, and is expressed by a truncated generalized polynomial chaos representation with $K$ basis functions:
\begin{equation}
V_{{\rm{out}}} (\vec \xi ,t) = \sum\limits_{k = 1}^{K} {v_k } (t)\psi _k (\vec \xi )\nonumber
\end{equation}
where $v_k(t)$ is the time-dependent coefficient of the generalized polynomial chaos expansion for the periodic steady-state solution and is actually solved at a set of time points during the entire period $[0, T]$. Next, $v_k(t)$ is expressed by a truncated Fourier series:
\begin{equation}
v_k (t) = \frac{a_k^0}{2}  + \sum\limits_{j = 1}^J {\left( {a_k^j \cos (j\omega t) + b_k^j \sin (j\omega t)} \right)} \nonumber
\end{equation}
with $\omega=\frac{2\pi}{T}$. The coefficients $a_k^j$ and $b_k^j$ 
\begin{equation}
a_k^j  = \frac{2}{T}\int\limits_0^T {v_k (t)\cos (j\omega t)dt,\;\;} b_k^j  = \frac{2}{T}\int\limits_0^T {v_k (t)\sin (j\omega t)dt\;} \nonumber
\end{equation}
are computed by a Trapezoidal integration along the time axis. Finally, the parameter-dependent total harmonic distortion is obtained as
\begin{equation}
\begin{array}{l}
 \hat x={\rm{THD}}  = f(\vec \xi ) = \sqrt {\frac{{\sum\limits_{j = 2}^J {\left[ {\left( {a^j (\vec \xi )} \right)^2  + \left( {b^j (\vec \xi )} \right)^2 } \right]} }}{{\left( {a^1 (\vec \xi )} \right)^2  + \left( {b^1 (\vec \xi )} \right)^2 }}}  \\ 
 {\rm{with}}\;a^j (\vec \xi ) = \sum\limits_{k = 1}^K {a_k^j \phi _k (\vec \xi )} ,\;\;b^j (\vec \xi ) = \sum\limits_{k = 1}^K {a_k^j \phi _k (\vec \xi )} . 
 \end{array}
\end{equation}
We set $J=5$ in the Fourier expansion, which is accurate enough for this low-noise amplifier. We use a 3rd-order generalized polynomial chaos expansion, leading to $K$$=$$35$. This surrogate model is evaluated by Monte Carlo with $5\times 10^5$ samples at the cost of $330$ seconds.

\textit{Density Estimation:} $114$ points are selected from the obtained cumulative density function curve to generate $p(x)$ and $\rho(x)$ by Alg. 1 and Alg. 2, respectively, which costs only several milliseconds. After scaling, Fig.~\ref{fig:LNA_density} shows the closed-form density functions for the total harmonic distortion of this low-noise amplifier, which matches the results from Monte Carlo simulation very well. The generated $p(x)$ monotonically increases from $0$ to $1$, and $\rho(x)$ is non-negative. Therefore, the obtained density functions are physically consistent.

\begin{figure}[t]
	\centering
		\includegraphics[width=3.3in]{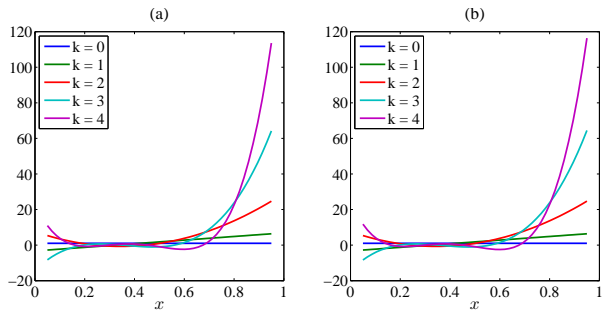} 
\caption{Computed generalized polynomial-chaos basis functions $\phi_k(x)$ ($k=0,\cdots,4$) for the low-noise amplifier. (a) uses the probability density function from Alg. 1, and (b) uses the probability density function from Alg. 2.}
	\label{fig:LNA_gpc}
\end{figure}

\textit{Basis Function:} Using the obtained density functions, several orthonormal polynomials of $x$ are constructed. Fig.~\ref{fig:LNA_gpc} shows the first five basis functions of $x$. Again, the basis functions resulting from our two density estimation implementations look similar since the density functions from both methods are only slightly different.

\textit{Gauss Quadrature Rule:} Five Gauss quadrature points and weights are computed and listed in Table~\ref{tab:Gauss_LNA}. Again the results from two density estimations are slightly different due to the employment of different density estimators.  When the density functions from piecewise cubic and piecewise rational quadratic interpolations are used, the the errors defined in (\ref{gpc_error}) are $3.11\times 10^{-14}$ and $4.34\times 10^{-14}$, respectively.
\begin{table}[t]
	\centering
	\caption{Computed Gauss quadrature points and weights for the low-noise amplifier.}	
	\label{tab:Gauss_LNA}
	\smallskip 
  \begin{threeparttable}
		\begin{tabular}{|c|c|c|c|}
	\hline
		\multicolumn{2}{|c|}{with $\rho(x)$ from Alg. 1}	& \multicolumn{2}{|c|}{with $\rho(x)$ from Alg. 2} \\ \hline
		$x^k$ & $w^k$ & $x^k$ & $w^k$ \\ \thickhline				
		0.131542 &0.056766 & 0.140381 & 0.073309\\  \hline		
		 0.251826 &0.442773& 0.261373 & 0.470691\\  \hline
		0.385311 &0.4432588 & 0.395704 & 0.400100\\  \hline
		0.550101 &0.066816 & 0.561873 & 0.055096\\  \hline
		0.785055 &0.001056 & 0.798122 & 0.000803\\  \hline
	\end{tabular} 	
\end{threeparttable}	
\end{table}

\subsection{Comparison with Asymptotic Probability Extraction}
\label{subsec:apexResults}
Finally we test our examples by the previous asymptotic probability extraction algorithm~\cite{xli2004:Apex,Apex}. Since our surrogate models are not in linear quadratic forms, we slightly modify asymptotic probability extraction: as done in~\cite{Jaskirat:2006} we use Monte Carlo to compute the statistical moments. All other procedures are exactly the same with those in~\cite{xli2004:Apex,Apex}.
\begin{figure*}[t]
	\centering
		\includegraphics[width=150mm]{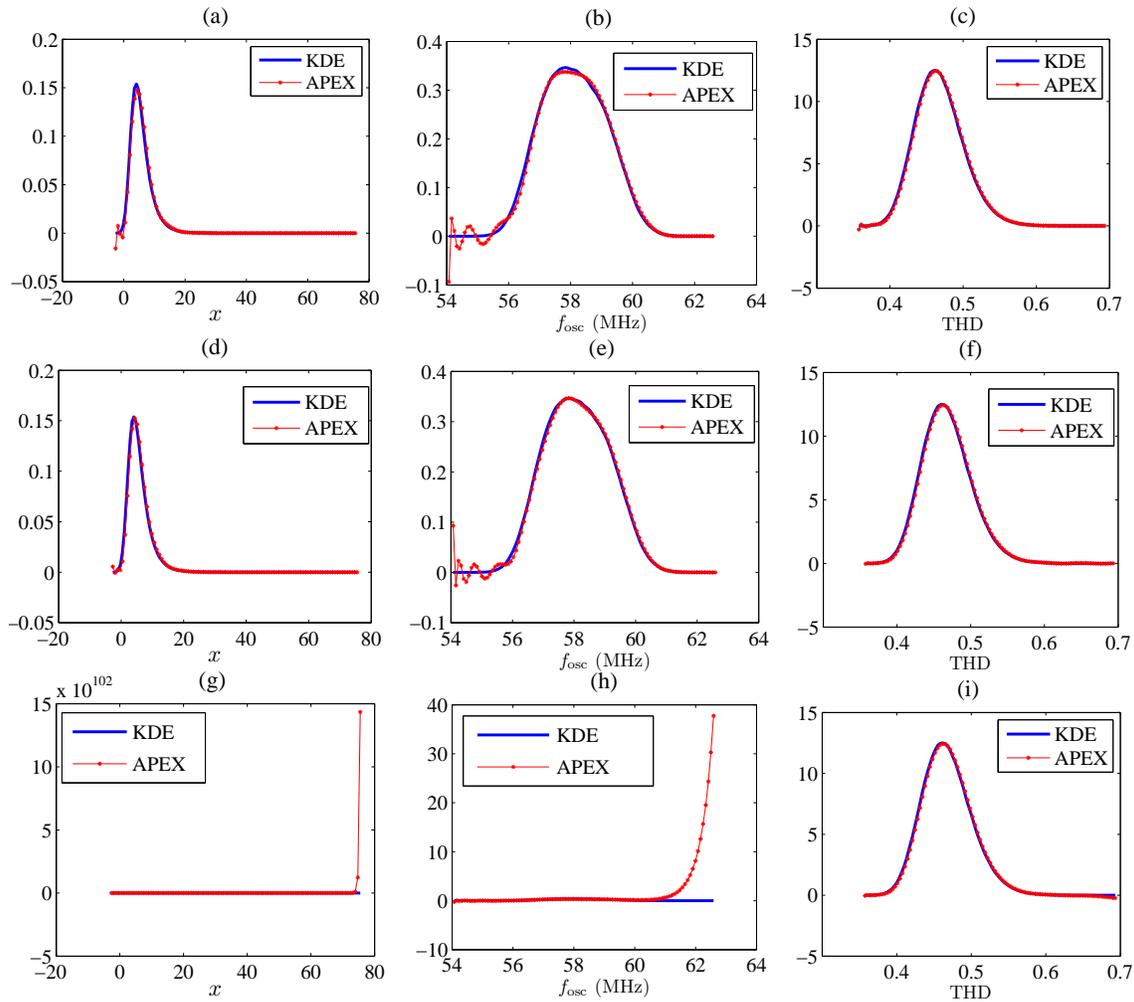} 
\caption{Probability density functions extracted by asymptotic probability extraction (APEX)~\cite{xli2004:Apex,Apex}, compared with the results from kernel density estimation (KDE). Left column: the synthetic example. Central column: frequency of the Colpitts oscillator. Right column:  total harmonic distortion (THD) of the low-noise amplifier. (a)-(c): with $10$ moments; (d)-(f): with $15$ moments; (g)-(i): with $17$ moments.}
	\label{fig:apex}
\end{figure*}

As shown in Fig.~\ref{fig:apex}, asymptotic probability extraction produces some negative probability density function values for the synthetic example and the Colpitts oscillator. The probability density functions of the low-noise amplifier are also slightly below $0$ in the tail regions, which is not clearly visible in the plots. Compared with the results from our proposed algorithms (that are non-negative and graphically indistinguishable with the original probability density functions), the results from asymptotic probability extraction have larger errors. As suggested by~\cite{xli2004:Apex,Apex}, we increase the order of moment matching to $15$, hoping to produce non-negative results. Unfortunately, Fig.~\ref{fig:apex} (d) and (e) show that negative probability density function values still appear, although the accuracy is improved around the peaks. Further increasing the order to $17$, we observe that some positive poles are generated by asymptotic waveform evaluation~\cite{AWE}. Such positive poles make the computed probability density functions unbounded and far from the original ones, as demonstrated by Fig.~\ref{fig:apex} (g) \& (h). For the low-noise amplifier, the approximated probability density function curve also becomes unbounded once we increase the order of moment matching to $20$, which is not shown in the plot. 

These undesirable phenomenon have been explained in Section~\ref{related_work} [c.f. Items $2)$ and $3)$]. Although it is possible to compute the statistical moments in some other ways (e.g., using maximum likelihood~\cite{cgu:2013} or point estimation method~\cite{PEM}), the shortcomings of asymptotic waveform evaluation (i.e., numerical instability and causing negative impulse response for a linear system) cannot be overcome. Because the density functions from asymptotic probability extraction may be physically inconsistent, they cannot be reused in a stochastic simulator (otherwise non-physical results may be obtained). Since the obtained probability density function is not guaranteed non-negative, the computed $\kappa_i$ in the three-term relation~(\ref{recurrence}) may become negative, whereas (\ref{int_cal}) implies that $\kappa_i$ should always be non-negative.

\section{Conclusions}

Motivated by hierarchical uncertainty quantification, this paper has proposed a framework to determine generalized polynomial-chaos basis functions and Gauss quadrature rules from surrogate models. Starting from a general surrogate model, closed-form density functions have been constructed by two monotone interpolation techniques. It has been shown that the obtained density functions are physically consistent: the cumulative density function is monotone and bounded by $0$ and $1$; the probability density function is guaranteed non-negative. Such properties are not guaranteed by existing moment-matching density estimators. By exploiting the special forms of our obtained probability density functions, generalized polynomial-chaos basis functions and Gauss quadrature rules have been easily determined, which can be used for higher-level stochastic simulation. The effectiveness of our proposed algorithms has been verified by several synthetic and practical circuit examples, showing excellent efficiency (at the cost of milliseconds) and accuracy (with errors
 around $10^{-14}$). The obtained generalized polynomial-chaos basis functions and Gauss quadrature points/weights allow standard stochastic spectral methods to efficiently handle surrogate models in a hierarchical simulator.

\bibliographystyle{IEEEtran}
\bibliography{REF}

\begin{IEEEbiography}[{\includegraphics[width=1in,height=1.25in,clip,keepaspectratio]{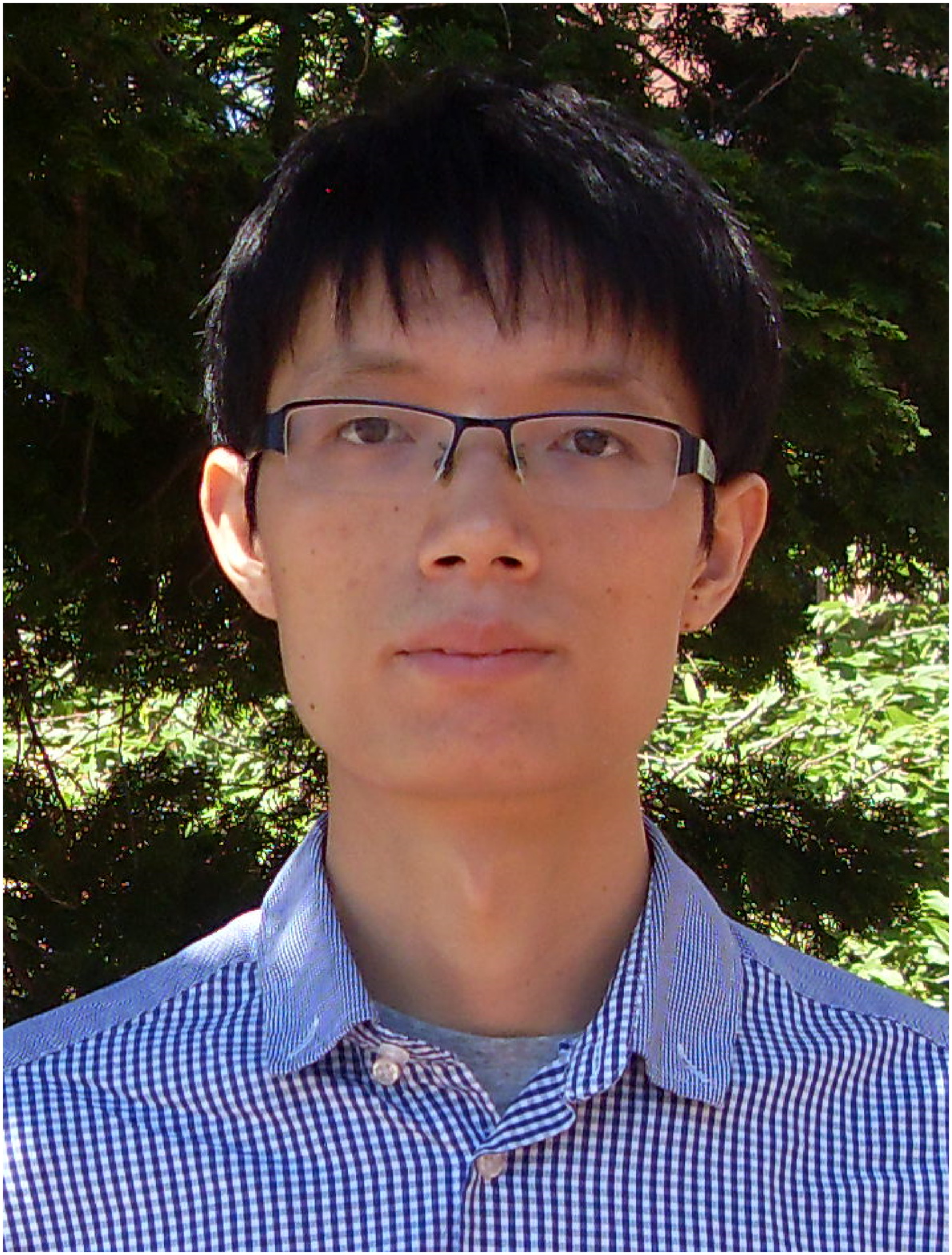}}]{Zheng Zhang} (S'09) received his B.Eng. degree from Huazhong University of Science and Technology, China, in 2008, and his M.Phil. degree from the University of Hong Kong, Hong Kong, in 2010. He is a Ph.D student in Electrical Engineering and Computer Science at the Massachusetts Institute of Technology (MIT), Cambridge, MA. His research interests include numerical methods for uncertainty quantification, computer-aided design (CAD) of integrated circuits and microelectromechanical systems (MEMS), and model order reduction.

In 2009, Mr. Zhang was a visiting scholar with the University of California, San Diego (UCSD), La Jolla, CA. In 2011, he collaborated with Coventor Inc., working on CAD tools for MEMS design. In the summer of 2013, he was a visiting scholar at the Applied Math Division of Brown University, Providence, RI. He was recipient of the Li Ka Shing Prize (university best M.Phil/Ph.D thesis award) from the University of Hong Kong, in 2011, and the Mathworks Fellowship from MIT, in 2010.
\end{IEEEbiography}

\begin{IEEEbiography}[{\includegraphics[width=1in,height=1.25in,clip,keepaspectratio]{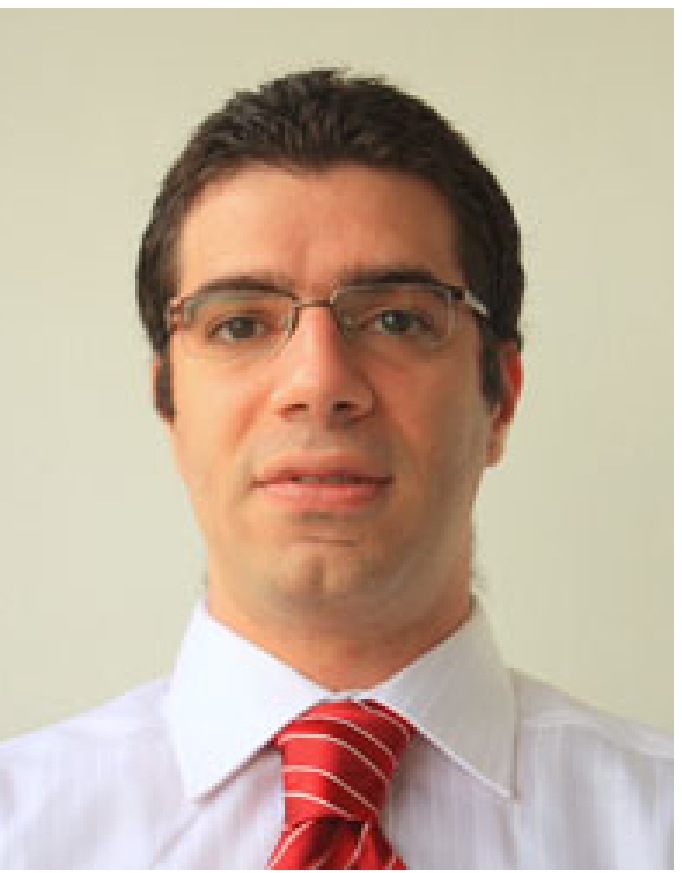}}]{Tarek El-Moselhy} received the B.Sc. degree in electrical engineering in 2000 and a diploma in mathematics in 2002, then the M.Sc. degree in mathematical engineering, in 2005, all from Cairo University, Cairo, Egypt. He received the Ph.D.
degree in electrical engineering from Massachusetts Institute of Technology, Cambridge, in 2010.

From 2010 to 2013, he was a postdoctoral associate in the Department of Aeronautics and Astronautics at Massachusetts Institute of Technology (MIT). Currently he is a Quantitative Analyst at The D. E. Shaw Group, New York, NY. His research interests include fast algorithms for deterministic and stochastic electromagnetic simulations, stochastic algorithms for uncertainty quantification in high dimensional systems, and stochastic inverse problems with emphasis on Bayesian inference. 

Dr. El-Moselhy received the Jin Au Kong Award for Outstanding PhD Thesis in Electrical Engineering from MIT in 2011, and the IBM Ph.D Fellowship in 2008.
\end{IEEEbiography}

\begin{IEEEbiography}[{\includegraphics[width=1in,height=1.25in,clip,keepaspectratio]{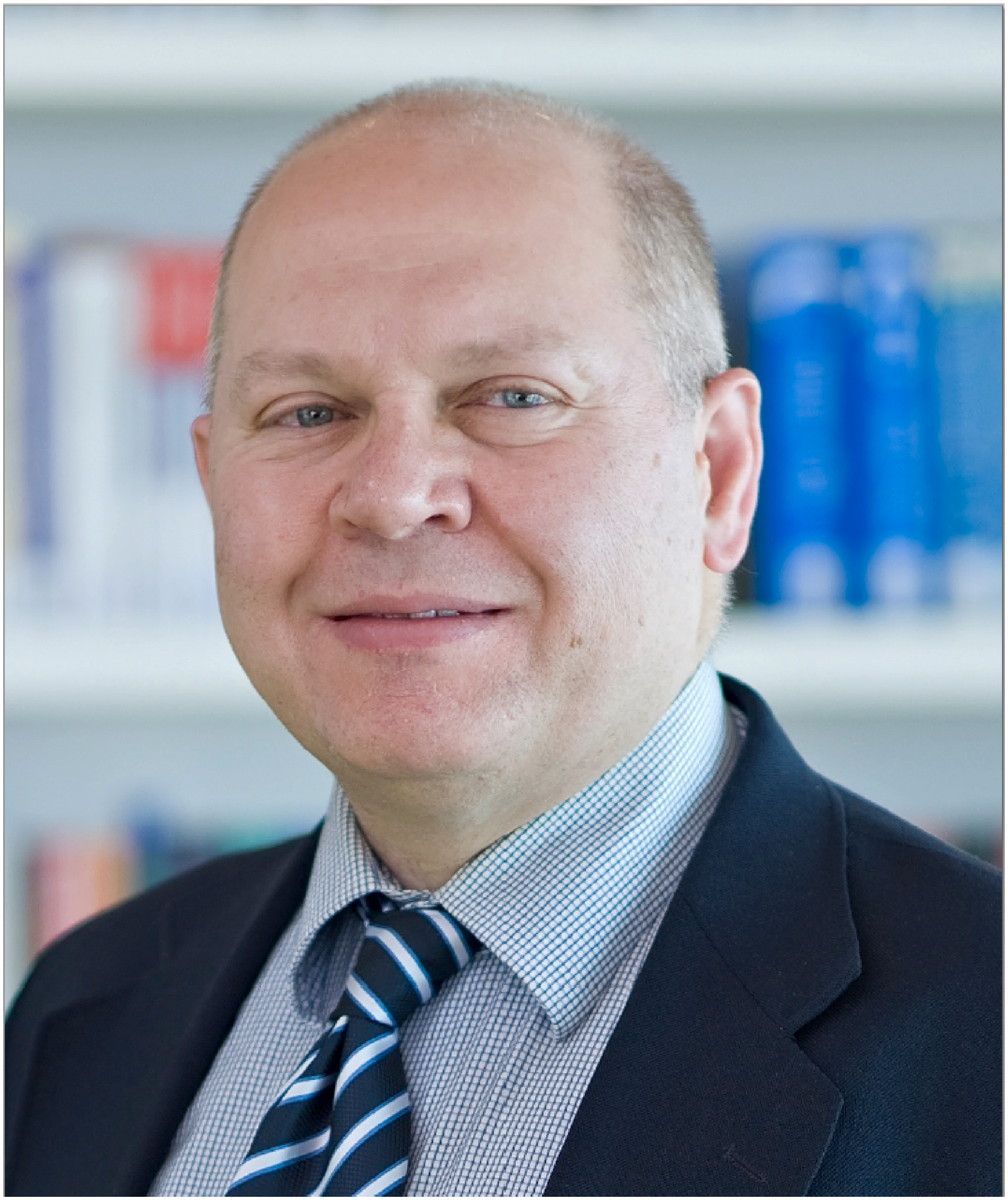}}]{Ibrahim (Abe) M. Elfadel} (SM'-02) received his Ph.D. from Massachusetts Institute of Technology (MIT) in 1993 and is currently Professor and Head of the Center for Microsystems at the Masdar Institute of Science and Technology, Abu Dhabi, UAE.

Dr. Elfadel is the Director of the TwinLab/Abu Dhabi Center for 3D IC Design, a joint R \& D program with the Technical University of Dresden, Germany. He is also the co-director of the ATIC-SRC Center of Excellence for Energy-Efficient Electronic Systems (ACE4S). Dr. Elfadel has 15 years of industrial experience with IBM in the research, development and deployment of advanced computer-aided design (CAD) tools and methodologies for deep-submicron, high-performance digital designs. His group's research is concerned with various aspects of energy-efficient digital system design and includes CAD for variation-aware, low-power nano-electronics, power and thermal management of multicore processors, embedded DSP for mmWave wireless systems, modeling and simulation of micro power sources, and 3D integration for energy-efficient VLSI design. 

Dr. Elfadel is the recipient of six Invention Achievement Awards, an Outstanding Technical Achievement Award and a Research Division Award, all from IBM, for his contributions in the area of VLSI CAD. He is currently serving as an Associate Editor for the IEEE Transactions on Computer-Aided Design for Integrated Circuits and Systems and the IEEE Transactions on Very-Large-Scale Integration.

\end{IEEEbiography}

\begin{IEEEbiography}[{\includegraphics[width=1in,height=1.25in,clip,keepaspectratio]{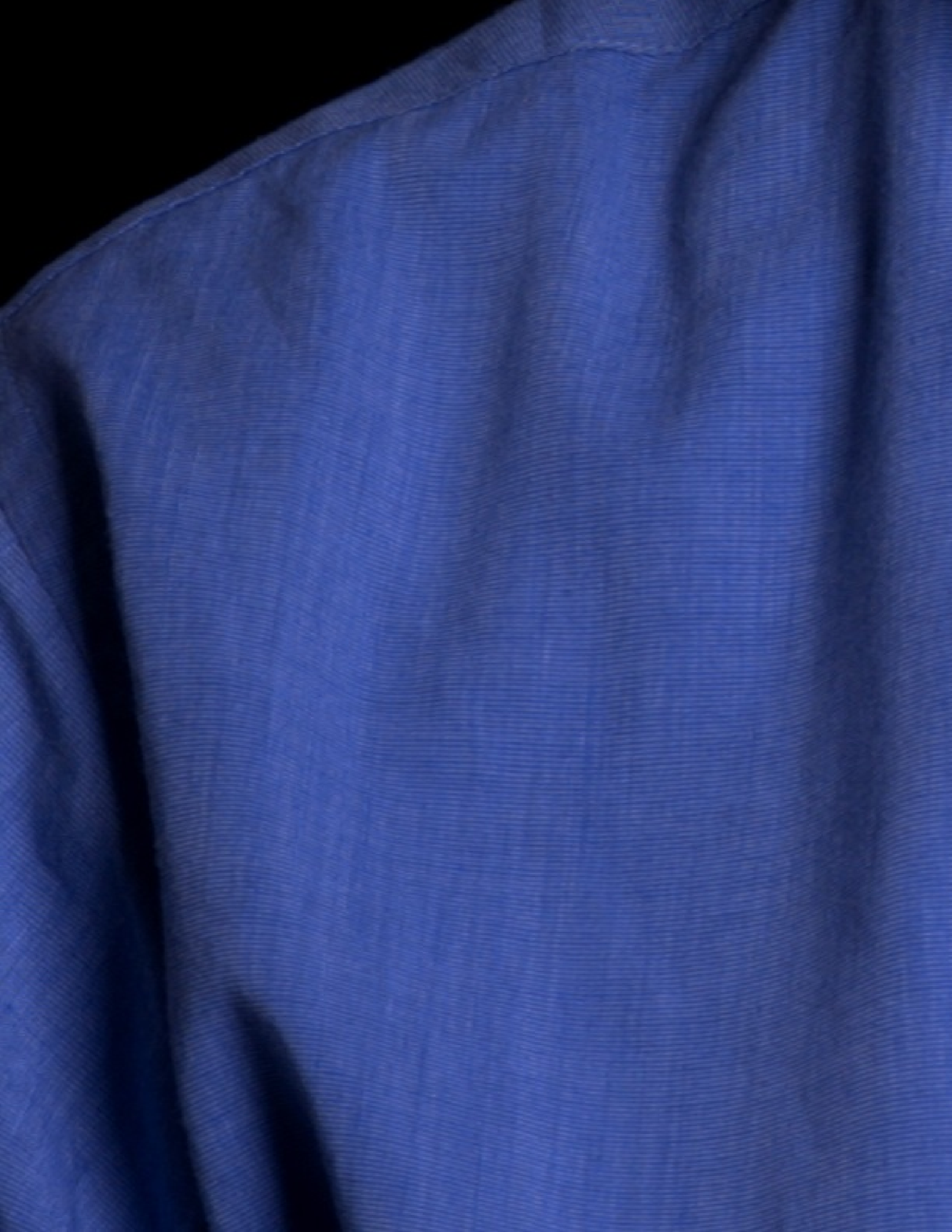}}]{Luca Daniel} (S'98-M'03) received the Laurea degree ({\it summa cum laude}) in electronic engineering from the Universita di Padova, Italy, in 1996, and the Ph.D. degree in electrical engineering from the University of California, Berkeley, in 2003.

He is an Associate Professor in the Electrical Engineering and Computer Science Department of the Massachusetts Institute of Technology (MIT), Cambridge. His research interests include accelerated integral equation solvers and parameterized stable compact dynamical modeling of linear and nonlinear dynamical systems with applications in mixed-signal/RF/mm-wave circuits, power electronics, MEMs, and the human cardiovascular system.

Dr. Daniel received the 1999 IEEE TRANSACTIONS ON POWER ELECTRONICS best paper award, the 2003 ACM Outstanding Ph.D. Dissertation Award in Electronic Design Automation, five best paper awards in international conferences, the 2009 IBM Corporation Faculty Award, and 2010 Early Career Award from the IEEE Council on Electronic Design Automation.
\end{IEEEbiography}

\end{document}